\documentstyle[12pt]{article}
\begin{document}
\begin{center}
{\large\bf THE FRACTIONAL KINETIC EQUATION \\
AND\\ 
THERMONUCLEAR FUNCTIONS}\par
\end{center}
\vspace{1cm}
\begin{center}
{\bf H.J. Haubold}\\
\end{center}
Outer Space Office, United Nations, Vienna International Centre,\\
P.O.Box 500, 1400 Vienna, Austria
\begin{center}
and\\[0.3cm]
{\bf A.M. Mathai}
\end{center}
Department of Mathematics and Statistics, McGill University,\\
805 Sherbrooke Street West, Montreal, Quebec, Canada H3A 2K6\par
\bigskip 
\noindent
{\bf Abstract.}
The paper discusses the solution of a simple kinetic equation of the type used for the computation of the change of the chemical composition in stars like the Sun. Starting from the standard form of the kinetic equation it is generalized to a fractional kinetic equation and its solutions in terms of H-functions are obtained. The role of thermonuclear functions, which are also represented in terms of G- and H-functions, in such a fractional kinetic equation is emphasized. Results contained in this paper are related to recent investigations of possible astrophysical solutions of the solar neutrino problem.
\clearpage
\noindent
\begin{center}
{\large\bf 1. Introduction}\par
\end{center}
\medskip
\noindent

A spherically symmetric, non-rotating, non-magnetic, self-gravitating model of a star like the Sun is assumed to be in thermal equilibrium and hydrostatic equilibrium, with a non-uniform chemical composition throughout. The star is characterized by its mass, luminosity, effective surface temperature, radius, central density, and central temperature. For a given mass, four of these variables are independent and are governed by four simultaneous, non-linear, ordinary differential equations of the first order and four boundary conditions. Since there are four equations but more than four unknowns, additional information must be provided through the equation of state, nuclear energy generation rate, and the opacity (constitutive equations). The assumptions of thermal equilibrium and hydrostatic equilibrium imply that there is no time dependence in the equations describing the internal structure of the star (Kourganoff, 1973, Perdang, 1976, Clayton, 1983).\par
The evolution of a star like the Sun is governed by a second system of differential equations, the kinetic equations, describing the rate of change of chemical composition of the star for each species in terms of the reaction rates for destruction and production of that species (Kourganoff, 1973, Perdang, 1976, Clayton, 1983).\par
Methods for modeling processes of destruction and production have been developed for bio-chemical reactions and their unstable equilibrium states (Murray, 1989) and for chemical reaction networks with unstable states, oscillations, and hysteresis (Nicolis and Prigogine, 1977). Stability investigations of thermonuclear reactions of stellar interest have not yet been worked out in detail. However, the potentiality of instabilities in thermonuclear chains may not be overlooked, since, as was pointed out once by Eddington, ``what is possible in the (Cavendish) Laboratory may not be too difficult in the Sun'' (Perdang, 1976, Mestel, 1999).\par
Consider an arbitrary reaction characterized by a time dependent quantity $N = N(t)$. It is possible to equate the rate of change $dN/dt$ to a balance between the destruction rate $d$ and the production rate $p$ of $N$, that is $dN/dt = -d + p$. In general, through feedback or other interaction mechanisms, destruction and production depend on the quantity $N$ itself: $d = d(N)$ or $p = p(N)$. This dependence is complicated since the destruction or production at time $t$ depends not only on $N(t)$ but also on the past history $N(\tau), \tau < t$, of the variable $N$. This may be formally represented by 
\begin{equation}
dN/dt = -d(N_t) + p(N_t),
\end{equation}
where $N_t$ denotes the function defined by $N_t(t^*) = N(t - t^*)$, $t^* > 0$. Here $d$ and $p$ are functionals and eq. (1) represents a functional-differential equation. In the following we study a special case of this equation, namely the equation
\begin{equation}
dN/dt = -\alpha N(t)
\end{equation}
with a constant $\alpha > 0$. Eq. (2) implies that spatial fluctuations or inhomogenieties in the quantity $N(t)$ are neglected. The standard solution of the differential equation (2) will be briefly discussed in Section 2 and the generalization to a fractional differential equation and its solution will be derived in Section 3. Conclusions will be drawn in Section 4.  
\bigskip
\noindent
\begin{center}
{\large\bf 2. Standard Kinetic Equation}\par
\end{center}
\medskip
\noindent
The production and destruction of species is described by kinetic equations governing the change of the number density $N_i$ of species $i$ over time, that is,
\begin{equation}
\frac{dN_i}{dt}=-\sum_jN_iN_j<\sigma\upsilon>_{ij}+\sum_{k,l\neq i}N_kN_l<\sigma\upsilon>_{kl},
\end{equation}
where $<\sigma\upsilon>_{mn}$ denotes the reaction probability for an interaction involving 
species $m$ and $n$, and the summation is taken over all reactions which either produce or destroy the species $i$ (Haubold and Mathai, 1995). For a gas of mass density $\rho$, the number density $N_i$ of the species $i$ is expressed in terms of its abundance $X_i$, by the relation $N_i=\rho N_AX_i/A_i,$ where $N_A$ is Avogadro's number and $A_i$ is the mass of $i$ in mass units. The mean lifetime $\tau_j(i)$ of species $i$ for destruction by species $j$ is given by the relation
\begin{equation} 
\lambda_j(i)=\frac{1}{\tau_j(i)}=N_j<\sigma\upsilon>_{ij}=\rho N_A\frac{X_j}{A_j}<\sigma\upsilon>_{ij},
\end{equation}
where $\lambda_j(i)$ is the decay rate of $i$ for interactions with $j$. Eq. (4) reveals the physical importance of $<\sigma\upsilon>_{ij}$ for the kinetic equation (3).\par
In the case of a nondegenerate, nonrelativistic gas, if a nonresonant charged nuclear reaction proceeds at low energies dominated by Coulomb-barrier penetration, the reaction probability $<\sigma\upsilon>_{mn}$ takes the form (Clayton, 1983, Bergstroem et al., 1999)

\begin{equation}
<\sigma\upsilon>_{mn}=(\frac{8}{\pi\mu})^{1/2}\sum^2_{\nu=0}\frac{1}{(kT)^{-\nu+1/2}}\frac{S^{(\nu)}(0)}{\nu!}I_2(\nu,a,z,\rho),
\end{equation}
where $I_2$ represents a thermonuclear function given by
\begin{eqnarray}
I_2(\nu - 1,a,z,\rho)& = & \int_0^\infty dyy^{\nu-1}e^{-ay-zy^{-\rho}}, \;\;\;\nu >0,a>0, z>0, \rho>0\nonumber\\
& = & \frac{a^{-\nu}}{\rho}H^{2,0}_{0,2}\left[az^{\frac{1}{\rho}}\left|_{(\nu,1),(0,\frac{1}{\rho})}\right.\right],\label{line2}
\end{eqnarray}
where $H$ denotes Fox's H-function (Mathai, 1993, Haubold and Mathai, 1998), which was introduced into mathematical analysis to unify and extend existing results on symmetrical Fourier kernels.\par
Fox's H-function (Mathai and Saxena, 1973, 1978) is defined in terms of a Mellin-Barnes type integral 
\begin{equation}
H(z)=H^{m,n}_{p,q}\left[z|^{(a_1,\alpha_1),\ldots,(a_p,\alpha_p)}_{(b_1,\beta_1),\ldots,(b_q,\beta_q)}\right]=\frac{1}{2\pi i}\int_L h(s)z^{-s}ds,
\end{equation}
where
\begin{equation}
h(s)=\frac{\{\Pi^m_{j=1}\Gamma(b_j+\beta_js)\}\{\Pi^n_{j=1}\Gamma(1-a_j-\alpha_js)\}}{\{\Pi^q_{j=m+1}\Gamma(1-b_j-\beta_js)\}\{\Pi^p_{j=n+1}\Gamma(a_j+\alpha_js)\}}
\end{equation}
$i=\sqrt{-1}$, and $L$ is a suitable path which will be briefly described in the following. An empty product is interpreted as unity and it is assumed that the poles of $\Gamma(b_j+\beta_js), j=1,\ldots,m,$ are separated from the poles of $\Gamma(1-a_j-\alpha_js), j=1,\ldots,n; a_1,\ldots,a_p, b_1,\ldots,b_q$ are complex numbers; $\alpha_1,\ldots, \alpha_p, \beta_1,\ldots,\beta_q$ are positive real numbers. The poles of $\Gamma(b_j+\beta_js), j=1,\ldots,m,$ are at the points
\[s=-(b_j+\nu)/\beta_j, \;\;\;j=1,\ldots,m, \nu=0,1,\ldots\]
and the poles of $\Gamma(1-a_j-\alpha_js), j=1,\ldots,n,$ are at
\[s=(1-a_k+\lambda)/\alpha_k, \;\;\;k=1,2,\ldots,n, \lambda=0,1,\ldots\]
The condition of separability of these two sets of poles imposes that there be a strip in the complex s-plane where the H-function has no poles. There are three types of paths $L$ possible. These correspond to paths 1,2, and 3 shown in Fig. 1.\par
\bigskip
\begin{center}
Fig. 1.
\end{center}
For all practical problems where H-functions are to be applied one mainly requires paths 2 and 3. Hence the conditions coming from these two paths are given in the following. It is to be pointed out that when more than one path $L$ makes sense then it can be shown that they lead to the same function and thus there will be no ambiguity.\\
Let
\begin{equation}
\mu=\sum_{j=1}^q \beta_j-\sum^p_{j=1}\alpha_j
\end{equation}
and
\begin{equation}
\beta=\Pi^p_{j=1}\alpha_j^{\alpha j}\Pi^q_{j=1}\beta_j^{-\beta_j}.
\end{equation}
The H-function exists for the following cases:\\[0.3cm]
\[
\begin{array}{ll}
\mbox{{\bf Case (i)}} & q \geq 1, \mu>0, H(z)\;\; \mbox{exists for all}\; z, z\neq 0.\\
\mbox{{\bf Case (ii)}} & q\geq 1, \mu=0, H(z)\;\; \mbox{exists for}\; |z|<\beta^{-1}.\\
\mbox{{\bf Case (iii)}} & p\geq 1, \mu<0, H(z)\;\; \mbox{exists for all}\; z, z\neq 0.\\
\mbox{{\bf Case (iv)}} & p\geq 1, \mu=0, H(z)\;\; \mbox{exists for}\; |z|>\beta^{-1}.\\
\end{array}
\]\\[0.3cm]
In the above cases it is assumed that the basic condition is satisfied that the poles of $\Gamma(b_j+\beta_js), j=1,\ldots,m$ and $\Gamma(1-a_j-\alpha_js), j=1,\ldots,n$ are separated.  Note that in cases (i) and (ii) the H-function is evaluated as the sum of the residues at the poles of $\Gamma(b_j+\beta_js), j=1,\ldots,m$ and in cases (iii) and (iv) the H-function is evaluated as the sum of the residues at the poles of $\Gamma(1-a_j-\alpha_js), j=1,\ldots,n$.\par
When $\alpha_1=\ldots = \alpha_p =\beta_1 =\ldots = \beta_q =1$, the H-function reduces to Meijers's G-function. That is,
\begin{equation}
H^{m,n}_{p,q}\left[z|^{(a_1,1),\ldots,(a_p,1)}_{(b_1,1),\ldots,(b_q,1)}\right]=G^{m,n}_{p,q}\left[z|^{a_1,\ldots,a_p}_{b_1,\ldots,b_q}\right].
\end{equation}\par
When $\rho$ in eq. (6) is real and rational, then the H-function can be reduced to a G-function by using the multiplication formula for gamma functions.
\begin{equation}
\Gamma(mz)=(2\pi)^{\frac{1-m}{2}}m^{mz-\frac{1}{2}}\Pi^{m-1}_{j=0}\Gamma\left(z+\frac{j}{m}\right), m=1,2,\ldots
\end{equation}
 In the case of Coulomb-barrier penetration (Gamow factor), $\rho=\frac{1}{2}, I_2$ reduces to 
\begin{equation}
I_2(\nu,a,z,\frac{1}{2})=\frac{a^{-(\nu+1)}}{\pi^{1/2}}G^{3,0}_{0,3}\left[\frac{az^2}{4}\left|_{\nu+1,a,\frac{1}{2}}\right.\right],
\end{equation}
where $G$ denotes Meijer's G-function, which was introduced into mathematical analysis in attempts to give meaning to the generalized hypergeometric function $_pF_q$ in the case $p > q+1$.\par
Proceeding with eq. (3), the first sum in eq. (3) can also be written as

\begin{equation}
-\sum_jN_iN_j<\sigma\upsilon>_{ij}=-N_i(\sum_jN_j<\sigma\upsilon>_{ij})=N_ia_i,
\end{equation}
where $a_i$ is the statistically expected number of reactions per unit volume per unit time destroying the species $i$. It is also a measure of the speed in which the reaction proceeds. In the following we are assuming that there are $N_j(j=1,\ldots, i,\ldots)$ species $j$ per unit volume and that for a fixed $N_i$ the number of other reacting species that interact with the i-th species is constant in a unit volume. Following the same argument we have for the second sum in eq. (3) accordingly,
\begin{equation}
+\sum_{k,l\neq i}N_kN_l<\sigma\upsilon>_{kl}=+N_ib_i,
\end{equation}
where $N_ib_i$ is the statistically expected number of the i-th species produced per unit volume per unit time for a fixed $N_i$. Note that the number density of species $i, N_i=N_i(t)$, is a function of time while the $<\sigma\upsilon>_{mn}$, containing the thermonuclear functions (see eqs. (5) and (6)), are assumed to depend only on the temperature of the gas but not on the time $t$ and number densities $N_i$. Then eq. (1) implies that
\begin{equation}
\frac{dN_i(t)}{dt}=-(a_i-b_i)N_i(t).
\end{equation}
For eq. (16) we have three distinct cases, $c_i=a_i-b_i>0, c_i<0,$
and $c_i=0$, of which the last case says that $N_i$ does not vary over time, which means that the forward and reverse reactions involving species $i$ are in equilibrium; such a value for $N_i$ is called a fixed point and corresponds to a steady-state behavior. The first two cases exhibit that either the destruction $(c_i>0)$ of species $i$ or production $(c_i<0)$ of species $i$ dominates.\par
\noindent
For the case $c_i>0$ we have
\begin{equation}
\frac{dN_i(t)}{dt}=-c_iN_i(t),
\end{equation}
with the initial condition that $N_i(t=0)=N_0$ is the number density of species $i$ at time $t=0$, and it follows that
\begin{equation}
N_i(t)dt=N_0e^{-c_it}dt.
\end{equation}\par
The exponential function in eq. (18) represents the solution of the linear one-dimensional differential equation (17) in which the rate of destruction of the variable is proportional to the value of the variable. Eq. (17) does not exhibit instabilities, oscillations, or chaotic dynamics, in striking contrast to its cousin, the logistic finite-difference equation (Perdang, 1976, Haubold and Mathai, 1995). A thorough discussion of eq. (17) and its standard solution in eq. (18) is given in Kourganoff (1973). 
\bigskip
\noindent
\begin{center}
{\large\bf 3. Fractional Kinetic Equation}\par
\end{center}
\medskip
\noindent
In the following, for the sake of brevity, the index $i$ in eq. (17) will be dropped. The standard kinetic equation (17) can be integrated
\begin{equation}
N(t)-N_0=-c\;_0D_t^{-1}N(t),
\end{equation}
where $_0D_t^{-1}$ is the standard Riemann integral operator. The generalization of this
operator to the fractional integral of order $p>0$ is denoted by $_aD^{-p}_t$ and is defined, following Riemann-Liouville, based on the Cauchy formula, by
\begin{equation}
_aD_t^{-p}f(t)=\frac{1}{\Gamma(p)}\int^t_a d\tau f(\tau) (t-\tau)^{p-1},\;\;\;p>0
\end{equation}
with
$$_aD_t^0f(t)=f(t)$$\\
(Oldham and Spanier, 1974, Miller and Ross, 1993). The most general fractional integral operator of the type (20) contains Fox's H-function (7) as the kernel function. If $f(t)$ is continuous for $t\geq a$, then integration of arbitrary real order has the  property
$$_aD_t^{-p}(_aD_t^{-q}f(t))=_aD_t^{-p-q}f(t).$$\\
Replacing the Riemann integral operator by the fractional Riemann-Liouville operator $_0D_t^{-\nu}$ in eq. (19), we obtain a fractional integral equation corresponding to eq. (17)
\begin{equation}
N(t)-N_0=-c^\nu\;_0D_t^{-\nu}N(t).
\end{equation}
For dimensional reasons, the coefficient $c$ in eq. (19), containing the probabilities of the reaction under consideration, had to be replaced by $c^\nu$ accordingly.\par
The Laplace transform of the Riemann-Liouville fractional integral is
\begin{equation}
L\left\{_0D_t^{-p}f(t);p\right\}=p^{-p}F(p),
\end{equation}
where
$$F(p)=\Gamma(p)\int_{\tau=0}^{\infty}d\tau e^{-p\tau}f(\tau).$$
In order to solve eq. (21), the integral equation is exposed to a Laplace transformation
leading to
\begin{equation}
N(p)=L\{N(t);p\} =N_0 \frac{p^{-1}}{1+(\frac{p}{c})^{-\nu}}.
\end{equation}\par
To arrive at a representation of eq. (23) in terms of Fox's H-function, Mathai and Saxena's (1978) result can be used,
\begin{eqnarray}
\frac{z^\beta}{1+az^\alpha} & = & a^{-\frac{\beta}{\alpha}} H^{1,1}_{1,1}\left[az^\alpha \left|^{(\frac{\beta}{\alpha},1)}_{(\frac{\beta}{\alpha},1)}\right.\right],\nonumber\\
N(p) & = & N_0\frac{1}{c} H^{1,1}_{1,1}\left[\left(\frac{c}{p}\right)^\nu\left|^{(\frac{1}{\nu},1)}_{(\frac{1}{\nu},1)}\right.\right].
\end{eqnarray}
To prepare eq.(24) for an inverse Laplace transform, the following two fundamental properties of an H-function can be used,
\begin{equation}
\frac{1}{k}H^{m,n}_{p,q}\left[z|^{(a_1,\alpha_1),\ldots,(a_p,\alpha_p)}_{(b_1,\beta_1),\ldots,(b_q,\beta_q)}\right]= H^{m,n}_{p,q}\left[z^k|^{(a_1,k\alpha_1),\ldots,(a_p,k\alpha_p)}_{(b_1,k\beta_1),\ldots,(b_q,k\beta_q)}\right],k>0,
\end{equation}
\begin{equation}
H^{m,n}_{p,q}\left[z|^{(a_1,\alpha_1),\ldots,(a_p,\alpha_p)}_{(b_1,\beta_1),\ldots,(b_q, \beta_q)}\right]=H^{n,m}_{q,p}\left[\frac{1}{z}|^{(1-b_1,\beta_1),\ldots,(1-b_q,\beta_q)}_{(1-a_1,\alpha_1),\ldots,(1-a_p,\alpha_p)}\right],
\end{equation}
leading to
\begin{equation}
N(p)=N_0\frac{1}{c\nu}H^{1,1}_{1,1}\left[\frac{p}{c}\left|^{(1-\frac{1}{\nu}, \frac{1}{\nu})}_{(1-\frac{1}{\nu},\frac{1}{\nu})}\right.\right],
\end{equation}
where the H-function is defined in eq. (7).\par
The Laplace transform of Fox's H-function (7) is given in terms of another H-function by
\begin{equation}
L\left\{H(z);p\right\}=\frac{1}{p}H^{n+1,m}_{q,p+1}\left[p\left|^{(1-b_q,\beta_q)}_{(1,1),(1-a_p,\alpha_p)}\right.\right] 
\end{equation}
for $0\leq \mu \leq 1$ in (9), and\\
\begin{equation}
L\left\{H(z);p\right\}= \frac{1}{p} H^{m,n+1}_{p+1,q}\left[\frac{1}{p}\left|^{(0,1), (a_p,\alpha_p)}_{(b_q, \beta_q)}\right. \right]
\end{equation}
for $\mu\geq 1$ in (9), respectively. \par
Further, having $H(p)$, the inverse Laplace transform of this H-function is given by
\begin{equation}
H(z)=L^{-1}\left\{H(p),z\right\}=\frac{1}{z}H^{n,m}_{q,p+1}\left[z\left|^{(1-b_q, \beta_q)}_{(1-a_p, \alpha_p),(1,1)}\right.\right]
\end{equation}
for $0\leq\mu\leq 1$ in (9)
and\\
\begin{equation}
H(z)=L^{-1}\left\{H(p),z\right\}=\frac{1}{z}H^{m,n}_{p+1,q}\left[z\left|^{(a_p,\alpha_p), (0,1)}_{(b_q, \beta_q)}\right.\right]
\end{equation}
for $\mu\geq 1$ in (9), respectively.\\
The above four Laplace transforms hold for\\
$\max_{1\leq j \leq n} \Re\left(\frac{a_p^{-1}}{\alpha_p}\right) < \min_{1\leq j \leq m} \Re\left(\frac{b_q}{\beta_q}\right).$ \\
Applying an inverse Laplace transform to the H-function in eq. (27) gives
\begin{equation}
N(t)=N_0\frac{1}{\nu}H^{1,1}_{1,2}\left[ct\left| ^{(0,\frac{1}{\nu})}_{(0,\frac{1}{\nu}),(0,1)}\right.\right],
\end{equation}
which is the solution of the fractional kinetic equation (21).
For the H-function in eq. (7) with (32), the following computable representation can be derived (Mathai and Saxena, 1978).
When the poles of $\Pi^m_{j=1} \Gamma(b_j-\beta_js)$ are simple, that is,
\[\beta_h(b_j+\lambda)\neq\beta_j(b_h+\nu)\]
for $j\neq h; j,h=1,\ldots, m; \lambda, \nu=0,1,2,\ldots.$ Then one obtains the following expansion for the H-function,
\begin{eqnarray}
H^{m,n}_{p,q}(z)&=&\sum^m_{h=1}\sum^\infty_{\nu=0}\frac{\left\{\Pi^m_{j=1,j\neq h}\Gamma\left\{b_j-\beta_j\frac{(b_h+\nu)}{\beta_h}\right\}\right\}}{\left\{\Pi^q_{j=m+1}\Gamma\left\{1-b_j+\beta_j\frac{(b_h+\nu)}{\beta_h}\right\}\right\}}\nonumber\\
& \times & \frac{\left\{\Pi^n_{j=1}\Gamma\left\{1-a_j+\alpha_j\frac{(b_h+\nu)}{\beta_h}\right\}\right\}} {\left\{\Pi^p_{j=n+1}\Gamma\left\{a_j-\alpha_j\frac{(b_h+\nu)}{\beta_h}\right\}\right\}}\frac{(-1)^\nu z^{(b_h+\nu)/\beta_h}}{(\nu)!\beta_h},
\end{eqnarray}
which exists for all $z\neq 0$ if $\mu > 0 $ and for $0<|z|<\beta^{-1} $ if $\mu=0$,
where $\mu$ and $\beta$ are given in eqs. (9) and (10).
Comparing (32) with eq. (33), one obtains the series expansion
\begin{equation}
N(t)=N_0\sum^\infty_{k=0}\frac{(-1)^k}{\Gamma(\nu k+1)}(ct)^{\nu k}.
\end{equation}
For $\nu=1$, the exponential solution of the standard kinetic equation (18) is recovered.\par
\bigskip
\noindent
\begin{center}
{\large\bf  4. Conclusions}
\end{center}
The gravitationally stabilized solar fusion reactor, because of its density and temperature, is not a weakly interacting or high temperature plasma. Recently, Kaniadakis et al. (1997) and Coraddu et al. (1998) have explored the possibility that the electrical microfield distribution influences the particle dynamics, collisions are not two-body processes and retain memory beyond single scattering events, and that the velocity correlation function has long-time memory arising from the coupling of collective and individual degrees of freedom. In this connection particle diffusion is a non-Markovian process (anomalous diffusion) and diffusion and frictional coefficients are energy dependent. Based on these results, they conclude that the equilibrium statistical distribution function differs from the Maxwell-Boltzmannian one and is governed by generalized Boltzmann-Gibbs statistics. \par
Even if the deviations from the Maxwell-Boltzmann distribution, that are compatible with the current knowledge of the solar core behavior, are small, they are sufficient to sensibly modify the sub-barrier particle reaction rates and subsequently solar neutrino fluxes. The above authors have also shown that the respective modifications of the reaction rates do not affect bulk properties of the gravitationally stabilized solar core such as sound of speed or hydrostatic equilibrium which depend on the mean values obtained by averaging over the Maxwell-Boltzmann distribution function (Degli'Innocenti et al. 1998). \par
Haubold and Mathai (1998) have derived general closed-form representations of particle reaction rates that are suitable to incorporate changes of the Maxwell-Boltzmann distribution function, mindfully taking into account the fact that the whole distribution has physical meaning, contrary to the case of, say, the Gaussian law of errors. In this paper we proceeded one step further in generalizing the standard kinetic equation (17) to a fractional kinetic equation (21) and derived solutions of a fractional kinetic equation that contains the particle reaction rate (or thermonuclear function) (5) as a time constant, and provided the analytic technique to further investigate possible modifications of the reaction rate through a kinetic equation. The Riemann-Liouville operator in the fractional kinetic equation introduces a convolution integral with slowly-decaying power-law kernel, which is typical for memory effects referred to in Kaniadakis et al. (1997) and Coraddu et al. (1998). This technique may open an avenue to accommodate changes in standard solar model core physics (Dar and Shaviv, 1999) as proposed by Shaviv and Shaviv (1999) and Schatzman (1999). \par
In the solution of the fractional kinetic equation (21), given in eqs. (32) and (34), the standard exponential decay is recovered for $\nu = 1$. However, eqs. (32) and (34), for $0 < \nu < 1$, show power law behavior for $t\rightarrow \infty$ and are constant (initial value $N_0$) for $t\rightarrow 0$. In the investigations in this paper, we used, as an example, the standard thermonuclear function $I_2 ^{(\infty)}$ = $I_1$; the same computations can be done for $I_2 ^{(d)}$, $I_3$, or $I_4$ with additional parameters showing modifications of the Maxwell-Boltzmann distribution function (Haubold and Mathai, 1998). \par
All analytic results in this paper have been achieved by the application of the theory of generalized hypergeometric functions, particularly Meijer's G-function and Fox's H-function, which seem to be the natural means for tackling the problems referred to above.\par
\clearpage
\begin{center}
{\bf Acknowledgments}
\end{center}
\noindent
The investigation of formation of structure in a non-equilibrium solar fusion plasma was discussed in-depth on 12 September 1984 with Hans-Juergen Treder at the Einstein Laboratory for Theoretical Physics in Caputh. Short-time instabilities in solar nuclear reaction kinetics have been explored briefly in a letter of Jean Perdang dated 8 January 1991. The corresponding author of this paper (HJH) is grateful to both colleagues and wishes to put this on record even at such a late date. The same author acknowledges inspiring discussions on non-Markovian processes in a solar nuclear plasma with Piero Quarati at the occasion of the sixth UN/ESA Workshop on Basic Space Science at the Max-Planck-Institute for Radioastronomy in Bonn, 9-13 September 1996.\par
\bigskip
\begin{center}
{\large\bf References}\par
\end{center}
\bigskip
\noindent
Bergstroem, L., Iguri, S., and Rubinstein, H.: 1999, Constraints on the\\
variation of the fine structure constant from big bang nucleosynthesis,\\
{\em Phys. Rev.} {\bf D60}, 045005-1 - 045005-9.
\\[4mm]
Clayton, D.D.: 1983, {\em Principles of Stellar Evolution and Nucleosynthesis}, Second Edition, The University of Chicago Press, Chicago and London.
\\[4mm]
Coraddu, M., Kaniadakis, G., Lavagno, A., Lissia, M., Mezzorani, G., and Quarati, P.: 1998, Thermal distributions in stellar plasma, nuclear reactions and solar neutrinos,\par
\noindent
http://xxx.lanl.gov/abs/nucl-th/981108.
\\[4mm]
Dar, A. and Shaviv, G.: 1999, The solar neutrino problem - an update, {\em Physics Reports} {\bf 311}, 115-141.
\\[4mm]
Degli'Innocenti, S., Fiorentini, G., Lissia, M., Quarati, P., and Ricci, B.: 1998, Helioseismology can test the Maxwell-Boltzmann distribution,\par
\noindent
http://xxx.lanl.gov/abs/astro-ph/9807078.
\\[4mm]
Haubold, H.J. and Mathai, A.M.: 1995, A heuristic remark on the periodic variation in the number of solar neutrinos detected on Earth, {\em Astrophys. Space Sci.} {\bf 228}, 113-134.
\\[4mm]
Haubold, H.J. and Mathai, A.M.: 1998, On thermonuclear reaction rates, {\em Astrophys. Space Sci.} {\bf 258}, 185-199.
\\[4mm]
Kaniadakis, G., Lavagno, A., Lissia, M., and Quarati, P.: 1998, Anomalous\\
diffusion modifies solar neutrino fluxes, {\em Physica} {\bf A261}, 359-373,\par
\noindent
http://xxx.lanl.gov/abs/astro-ph/9710173.
\\[4mm]
Kourganoff, V.: 1973, {\em Introduction to the Physics of Stellar Interiors}, D. Reidel Publishing Company, Dordrecht.
\\[4mm]
Mathai, A.M.: 1993, {\em A Handbook of Generalized Special Functions for\\
Statistical and Physical Sciences}, Clarendon Press, Oxford.
\\[4mm]
Mathai, A.M. and Saxena, R.K.: 1973, {\em Generalized Hypergeometric\\
Functions with Applications in Statistics and Physical Sciences}, Springer-Verlag, Lecture Notes in Mathematics Vol. 348, Berlin Heidelberg New York.
\\[4mm]
Mathai, A.M. and Saxena, R.K.: 1978, {\em The H-function with Applications in Statistics and Other Disciplines}, John Wiley and Sons, New Delhi.
\\[4mm]
Mestel, L.: 1999, The early days of solar structure theory, {\em Physics Reports} {\bf 311}, 295-305.
\\[4mm]
Miller, K.S. and Ross, B.: 1993, {\em An Introduction to the Fractional Calculus and Fractional Differential Equations}, John Wiley and Sons, New York.
\\[4mm]
Murray, J.D.: 1989, {\em Mathematical Biology}, Biomathematics Texts Vol. 19, Springer-Verlag, Berlin.
\\[4mm]
Nicolis, G. and Prigogine, I.: 1977, {\em Self-Organization in\\
Nonequilibrium Systems - From Dissipative Structures to Order Through Fluctuations}, John Wiley and Sons, New York.
\\[4mm]
Oldham, K.B. and Spanier, J.: 1974, {\em The Fractional Calculus}, Academic Press, New York.
\\[4mm]
Perdang, J.: 1976, {\em Lecture Notes in Stellar Stability}, Part I and II, Instituto di Astronomia, Padova.
\\[4mm]
Schatzman, E.: 1999, Role of gravity waves in the solar neutrino problem, {\em Physics Reports} {\bf 311}, 143-150.
\\[4mm]
Shaviv, G. and Shaviv, N.J.: 1999, Is there a dynamic effect in the screening of nuclear reactions in stellar plasma?, {\em Physics Reports} {\bf 311}, 99-114.
\end{document}